# Hybrid optical-thermal devices and materials for light manipulation and radiative cooling

Svetlana V. Boriskina*, Jonathan K. Tong, Wei-Chun Hsu, Lee Weinstein, Xiaopeng Huang, James Loomis, Yanfei Xu, Gang Chen*

Department of Mechanical Engineering, Massachusetts Institute of Technology
Cambridge, MA 02139, USA

## ABSTRACT

We report on optical design and applications of hybrid meso-scale devices and materials that combine optical and thermal management functionalities owing to their tailored resonant interaction with light in visible and infrared frequency bands. We outline a general approach to designing such materials, and discuss two specific applications in detail. One example is a hybrid optical-thermal antenna with sub-wavelength light focusing, which simultaneously enables *intensity enhancement* at the operating wavelength in the visible *and reduction of the operating temperature*. The enhancement is achieved via light recycling in the form of whispering-gallery modes trapped in an optical microcavity, while cooling functionality is realized via a combination of reduced optical absorption and radiative cooling. The other example is a fabric that is *opaque in the visible range yet highly transparent in the infrared*, which allows the human body to efficiently shed energy in the form of thermal emission. Such fabrics can find numerous applications for personal thermal management and for buildings energy efficiency improvement.

## 1. INTRODUCTION

Modification of absorptive and radiative properties of materials beyond their classical limits is of fundamental interest and can benefit many applications, including solar energy conversion, nanoscale imaging and sensing, solid-state lighting, personal comfort technologies just to name a few. Many applications also require strikingly different response of materials to visible and infrared light[1–5]. This can be achieved by optimal combination of material properties and photon confinement effects in meso-scale structures tailored to either interact resonantly or not to interact at all with high- and low-energy photons.

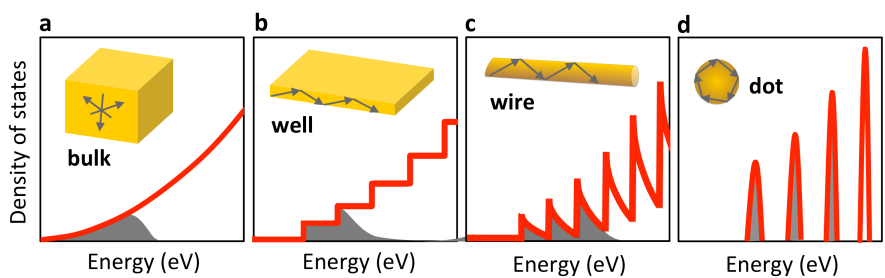

Figure 1. Spatial confinement concentrates photon states, which enhances optical absorption and potentially thermal emission at select frequencies. Red lines: density of optical states in low-dimensional emitters; Shaded areas: thermal distribution of photons in (a) bulk material, (b) wells, (c) wires, and (d) dots at a chosen temperature.

In particular, light focusing, scattering and thermal emission can be manipulated by confining photons in low-dimensional potential traps, in direct analogy with confining elections to quantum wells, wires and dots (Fig. 1). The thermal emission signal is defined by the available density of photon states and by the Bose-Einstein distribution of photons as a function of temperature. In general, the electromagnetic energy density of radiation in a material per unit frequency and angle is defined as[6,7]:

$$U = \hbar\omega \cdot D(\omega) \cdot (\exp(\hbar\omega/k_B T) - 1)^{-1} d\omega, \quad (1)$$

where $k_B$ is the Boltzmann constant, $h = 2\pi\hbar$ is the Planck's constant, $\omega$ is the photon angular frequency, and $T$ is the temperature of the material. $D(\omega)$ is the density of the photon states (DOS) defined per unit area per unit angle per unit frequency as follows:

$$D(\omega) = d\Omega/(2\pi)^3 \cdot k^2(\omega) \cdot dk/d\omega. \quad (2)$$

*sborisk@mit.edu; gchen2@mit.edu; http://www.mit.edu/~sborisk/; http://web.mit.edu/nanoengineering/

For plane waves of both polarizations in isotropic bulk dielectric with refractive index $n$ and for the full angular range of the emission $\Omega = 4\pi$, the density of states takes the familiar form that appears in the classical Planck's law of the thermal emission: $D = \omega^2 n^3/\pi^2 c^3$. However, as Eq. (1) suggests, thermal emission spectrum can be tailored not just by the temperature – as typical for the blackbody radiation – but also by the photon DOS, and this modification can be governed by strong optical confinement effects.

We have previously used the analogy between optical and quantum confinement to design a 'thermal well' thermophotovoltaic energy converter with both the emitter and the absorber in the form of ultra-thin semiconductor films. The efficiency of such a converter was predicted to exceed by over an order of magnitude both the bulk limit and the Shockley Queisser limit for a blackbody emitter[4]. By using the same principle, objects with the dimensions on or below the scale of the wavelength can be designed to provide strong resonant absorption of light by recycling incoming photons in the form of trapped optical modes with quantized eigenfrequency values[8] (Fig. 2a). Light absorption and emission can also be modified by surface phonon-[9] or plasmon-polariton modes[10,11], especially in combination with strong spatial confinement effects (Fig. 2b).

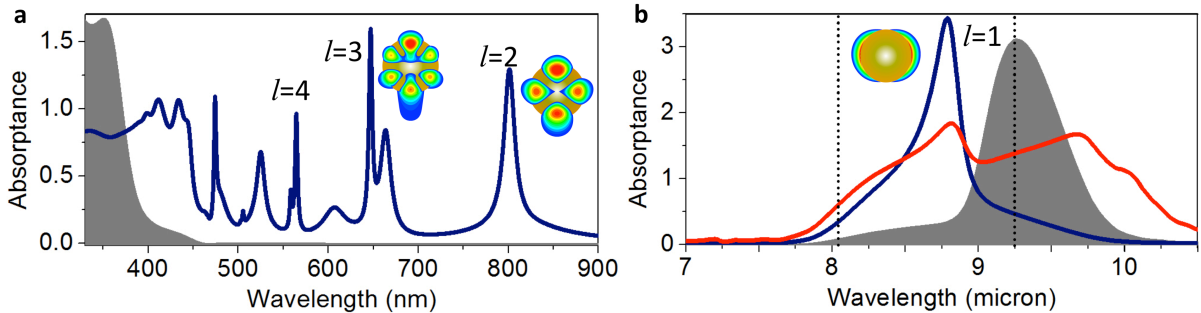

Figure 2. Spectral absorption/emission cross-sections of (a) a 200-nm-diameter Si nanosphere supporting trapped optical modes (*l* is the mode azimuthal momentum quantum numbers) and (b) SiO$_2$ microspheres supporting surface phonon polariton modes (navy: 1-micron-diameter sphere, red: 4-micron-diameter sphere). The plotted values are normalized to the geometrical cross-sections of the corresponding particles. The grey shaded areas illustrate the imaginary part of the particle refractive index (not to scale) as a function of photon wavelength. It can be clearly seen that the peaks of highest emittance/absorptance do not correlate with the regions of the highest material losses, and are instead governed by the optical confinement effects. Some peaks correspond to the modes with absorption cross-sections exceeding the particles geometrical cross-sections (the ratio larger than 1).

In the applications discussed in this paper, the goal is to make use of the above resonant effects to design hybrid devices and materials with drastically different electromagnetic response in the visible and infrared frequency ranges, for example, to design materials that are *efficient thermal emitters yet poor absorbers of incoming external radiation* (either solar or laser light). This objective *does not contradict the Kirchhoff's law*[12], whose common corollary is typically assumed to be that a good absorber has to be a good emitter, and a poor absorber - a poor emitter. However, what the Kirchhoff's law actually states is that "for an arbitrary body emitting & absorbing thermal radiation *in thermodynamic equilibrium*, the emissivity is equal to the absorptivity." However, the hybrid components and materials discussed in this work a*re not in thermal equilibrium with external sources* (i.e., the sun or a coherent laser source). They, however, still need to obey the principle of detailed balance, which leads to the general form of the Kirchhoff's law stating that

$$\varepsilon(\omega,\theta,\varphi) = \alpha(\omega,\theta,\varphi), \qquad (3)$$

where $\varepsilon(\omega,\theta,\varphi)$ is the directional spectral emittance, $\alpha(\omega,\theta,\varphi)$ is the directional spectral absorptance, $\omega$ is the frequency of the photon, and the angles $\theta,\varphi$ specify a direction of the photon propagation.

However, for a good absorber of external radiation at a frequency $\omega$ to become a good thermal emitter at the same frequency, it needs to be heated to a high enough temperature to fill the available photon states (illustrated as gray shaded areas in Fig. 1). On the other hand, a low-dimensional emitter can be designed to have resonantly-high photon DOS at select frequencies. If this DOS exceeds the corresponding value for the isotropic bulk dielectric, another common corollary of the Planck's and Kirchhoff's laws is no longer valid, namely, that the emissivity cannot exceed one, which is the highest value only attainable by a blackbody emitter. However, this corollary is valid only when 'the linear dimension of all parts considered, as well as radii of curvature of all surfaces... are large compared with the wavelength of the ray considered,'[6] and can be broken in the case of low-dimensional emitters with absorption cross-sections exceeding their geometrical cross-sections[13] (such as those in Figs. 2a,b). By using the above approach, we designed hybrid metal-dielectric nano-antennas that offer focusing enhancement by two orders of magnitude and simultaneous reduction of operating temperature by several hundred degrees with respect to their all-metal counterparts.

By using the same principles, low-dimensional objects can be tailored to strongly interact with short-wavelength visible light, yet not obstruct propagation of longer-wavelength infrared radiation. In particular, if the thickness of thin fibers (i.e., photon wires) is adjusted such that they are sub-wavelength for the infrared radiation, yet wavelength-scale for the visible one, they will be transparent for IR yet opaque for visible photons. We will discuss in the following how this strategy can be used to design new types of fabrics, which can provide personalized cooling by allowing thermal emission from the skin to pass through the clothes[5].

## 2. HYBRID OPTICAL-THERMAL ANTENNAS

Metallic nanoparticles supporting localized surface plasmon resonances have become an indispensable tool in sensing and nanoscale imaging applications. However, the high plasmon-enhanced electric field intensity in the visible and near-IR range that enables the above applications also causes excessive heating of metal nanoparticles, which is a major drawback in plasmonics.

Furthermore, ultrasensitive optical detection schemes, which benefit from high field concentration in plasmonics, also require high spectral resolution to reach single-molecule level of detection. High spectral resolution is typically unattainable in purely plasmonic sub-wavelength structures due to high dissipative losses of metals. We have previously demonstrated that this limitation can be overcome by combining plasmonic and high-Q photonic elements into hybrid optoplasmonic sensor platforms[14–19]. Among the emergent properties of optoplasmonic structures is their ability to simultaneously achieve extreme spectral and spatial localization of light. This is illustrated in Fig. 3a, which compares the field intensity enhancement in the plasmonic hot spot formed on a 150-nm diameter gold nanoparticle, which is either standalone (gray line) or a part of a hybrid metal-dielectric antenna with a 3-micron-diameter $SiO_2$ microsphere (red line). Over two orders of magnitude intensity enhancement can be observed at frequencies corresponding to the excitation of whispering gallery modes in the silica microsphere.

Our rigorous Mie-theory-based calculations also reveal that counter-intuitively, the field intensity enhancement in hybrid optoplasmonic structures may be accompanied by the *reduced* absorption in their metal constituents in comparison to the standalone nanoparticle (Fig. 3b). The reduced light absorption by itself should result in the reduction of the temperature of the antenna that will be reached under the steady-state laser illumination. We also show that, additionally, the infrared thermal emittance of a hybrid antenna can be tailored via the proper choice of both material and morphology of the dielectric elements having dimensions on the scale at or below the thermal emission peak wavelength. In particular, $SiO_2$ microsphere supports localized surface phonon polariton modes in the mid-to-far infrared spectral range, which significantly enhance the infrared thermal emittance of the hybrid meso-scale antenna over that of a metal nanoparticle (Figs. 2b and 3c)[20].

As a result, these hybrid systems can provide strong cooling of metal via the radiative heat extraction based on enhanced thermal emittance of the dielectric constituent (Fig. 3c). Overall, a combination of the strong light localization and enhancement achievable under lower operating temperatures in optoplasmonic materials is expected to yield a wide range of applications in plasmon-enhanced spectroscopy, sensing and imaging.

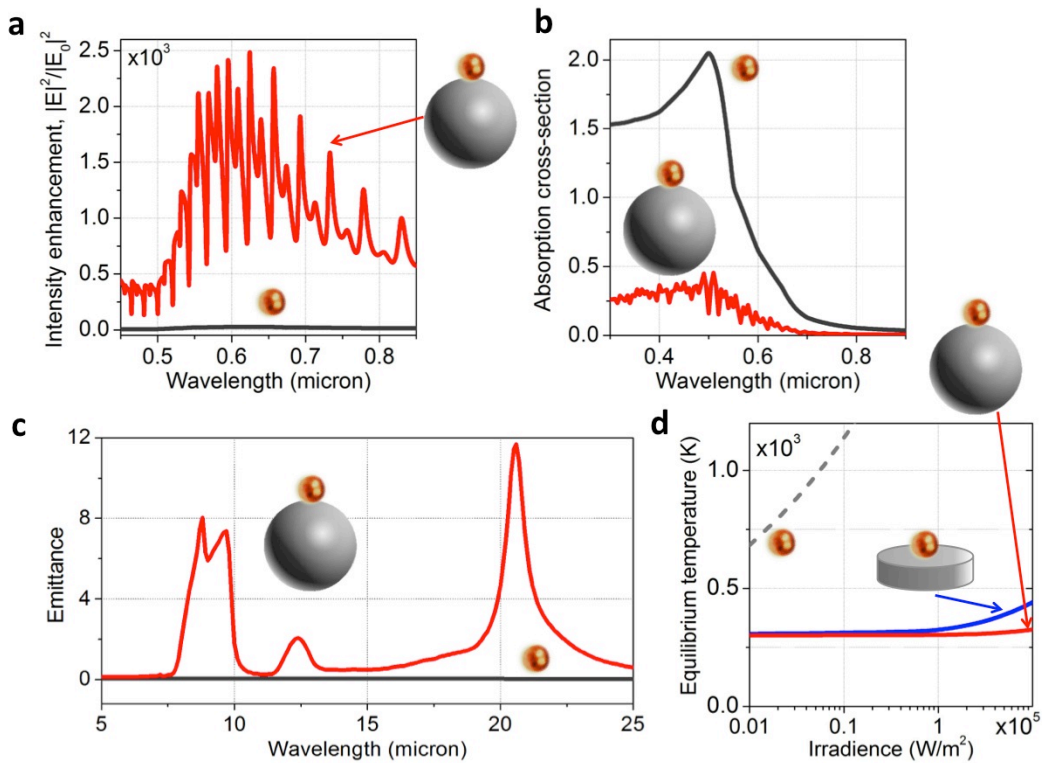

Figure 3. (a) Frequency spectrum of the local intensity enhancement on a surface of the 150nm-diameter Au nanoparticle, which is incorporated as a part of the optical-thermal antenna with a 3-micron-diameter dielectric microsphere. The sharp intensity peaks correspond to the excitation of WG modes in the dielectric sphere. The field intensity generated by a plane wave illuminating a standalone Au particle is shown as a gray line for a reference. (b) Absorption cross-sections of the hybrid and plasmonic nanoantennas with the same parameters as in (a) as a function of wavelength. (c) Frequency spectrum of the infrared thermal emittance of the hybrid antenna. The emittance spectrum of a standalone Au nanoparticle is shown for comparison as the gray line. Effective emittances normalized to the equivalent surface area are plotted (see Eq. 4). (d) Equilibrium temperature of the nanoparticle reached under steady-state illumination by a monochromatic plane wave with varying photon flux and with a frequency centered at the antenna highest intensity peak (solid red line: hybrid antenna; dashed gray line: standalone nanoparticle; solid blue line: nanoparticle on top of a planar silica surface of the same geometrical cross-section as the silica sphere).

## 3. VISIBLE-OPAQUE INFRARED-TRANSPARENT FABRICS

Personal thermo-regulation technologies aim to make people feel thermally comfortable in extreme environmental conditions while not limiting their mobility. Various technological solutions for personal cooling already are on the market. However, they are mostly tailored for athletes, the military, or first-responders, and are not very practical for everyday use. Many of the so-called active technologies – that need electrical power to operate – tend to be bulky and expensive. The passive personal cooling technology that thus far proven to be most successful is moisture wicking, where perspiration is drawn to the outer surface of the fabric and evaporated. Although it offers high mobility and ease of use, this technology relies on perspiration to provide cooling. While very useful for athletes, it is far from an optimum solution for sedentary individuals such as e.g., office workers.

By utilizing optical confinement effects in the polymer fibers, we have developed a simple solution to achieve passive cooling without perspiration. In particular, we have designed new types of fabrics, which can help people feel cooler by simply allowing thermal emission from the skin to pass through the clothes rather than to get trapped inside[5]. The new wearable technology is based on exploiting different mechanisms of electromagnetic waves scattering on obstacles that are either smaller or larger than the wavelength of the propagating field. Just like the sky is blue because the small air molecules scatter shorter-wavelength blue light more efficiently than longer-wavelength red light, the designed cooling fabrics are opaque for the visible light yet transparent for the long-wavelength thermal radiation from the human body (Fig. 4a). This property sets apart the proposed new fabrics from conventional ones, which mostly block thermal radiation emitted by the skin. The fabrics are optically designed by modeling photons interaction with single fibers, fiber bundles, and bundle arrays, which effectively act as flexible polymer optical-thermal metamaterials (Fig. 4b).

We developed a heat exchange model[5] to determine the required infrared optical properties of the new fabrics to ensure thermal comfort is maintained for environmental temperatures exceeding the normal room temperature. We experimentally observed that existing textiles fail to meet these requirements due to a combination of intrinsic material absorption and structural backscattering in the IR wavelength range (Fig. 4c).

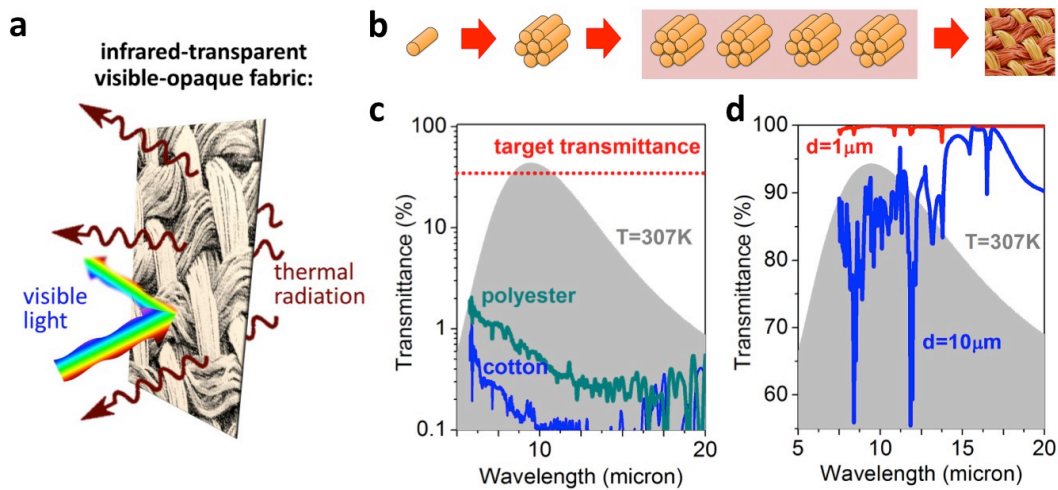

Figure 4. (a) Infrared-transparent visible-opaque fabric concept: while being opaque for visible light, the fabric allows thermal emission from the skin to penetrate through and dissipate in the environment. (b) Schematic of the fabric design strategy. (c) FTIR measurements of the IR transmittance of conventional fabrics (blue: cotton, teal: polyester). Red dotted line is the transmittance target value obtained from the heat transfer model. (d) Mie-theory calculations of the transmittance of fiber bundle arrays made of polyethylene fibers of varying diameters. Gray shaded areas are the blackbody emission spectra at T=307K (not to scale).

Designs of the new fabrics were developed using a combination of optimal material composition and structural photonic engineering. Specifically, synthetic polymers that support few vibrational modes were identified as candidate materials to reduce intrinsic material absorption in the IR wavelength range. To reduce backscattering losses, individual fibers were designed to be comparable in size to visible wavelengths in order to minimize reflection in the IR by virtue of weak Rayleigh scattering while remaining optically opaque in the visible wavelength range due to strong Mie scattering. By additionally reducing the size of the yarn, which is defined as a collection of fibers, less material is used thus decreasing volumetric absorption in the IR wavelength range even further. The new fabrics design was numerically demonstrated to exhibit high transparency in the IR wavelength range while remaining optically opaque in the visible wavelength range (Fig. 4d). Compared to conventional personal cooling technologies, these fabrics can provide fully passive means to cool the human body regardless of the person's physical activity level. However, additional studies and optimization of the level of comfort provided by such materials are needed.

The anticipated effect of the new optical-thermal polymer fabrics (tailored into a short sleeve shirt and pants) will be the increase in the body cooling rate by at least 23 W, which in turn will allow for raising summer HVAC set points by >4°F (to ~79°F) leading to significant energy savings.

## 4. CONCLUSIONS

We demonstrated the possibility to simultaneously manipulate optical scattering and absorption as well as thermal emission by confining photons in low-dimensional potential traps, in direct analogy with confining elections to quantum wells, wires and dots. By combining several materials supporting trapped and surface polariton modes, we designed composite nanostructures with tailored broadband spectral characteristics. In particular, we demonstrated hybrid metal-dielectric antennas that offer laser light focusing enhancement by two orders of magnitude and simultaneous reduction of operating temperature by several hundred degrees with respect to their all-metal counterparts[20].

In turn, thin fibers can be designed to absorb and scatter short-wavelength visible light, yet not obstruct propagation of longer-wavelength infrared radiation. We used this effect to propose new types of fabrics, which can provide personalized cooling by allowing thermal emission from the skin to pass through the clothes[5]. Simultaneously, the visible wavelength optical properties of the new fabrics are comparable to conventional textiles ensuring they are opaque to the human eye. This offers a novel, yet simplistic mechanism to provide personal cooling for everyday use[21].

## 5. ACKNOWLEDGEMENT

This work was supported by the U.S. Department of Energy, Office of Basic Energy Sciences, Division of Materials Science and Engineering Award No. DE-FG02-02ER45977.